\documentstyle[prb, aps, multicol]{revtex}

\def\etal{\emph{et al.}\ }

\input BoxedEPS
\SetepsfEPSFSpecial
\HideDisplacementBoxes

\begin{document}

\title{Low temperature thermal properties of mesoscopic
normal-metal/superconductor heterostructures}
\author{D.\ A.\ Dikin, S.\ Jung and V.\ Chandrasekhar}
\address{ Department of Physics and Astronomy, 
Northwestern University, Evanston, IL 60208, USA \\ }

\maketitle

\begin{abstract}Although the electrical transport properties of mesoscopic 
metallic samples have been investigated extensively over the past 
two decades, the thermal properties have
received far less attention.  This may be due in part to the difficulty of
performing thermal measurements on sub-micron scale samples.  We
report here quantitative measurements of the thermal conductance
and thermopower of a hybrid normal-metal/superconductor
heterostructure, which are made possible by the recent development
of a local-thermometry technique.  As with electrical transport
measurements, these thermal measurements reveal signatures of the
phase coherent nature of electron transport in these devices.
\end{abstract}


\begin{multicols}{2}

As the packing density of electronic devices on a single chip
continues to increase, the issue of heat transport and dissipation
in nanometer scale structures becomes of increasing importance.
Although many experiments have focused on the electrical
properties of micro- and nanometer scale structures, the thermal
characteristics of such devices are only beginning to be explored.
In addition to addressing critical issues related to the
fabrication of the next generation of electronic devices,
exploration of the thermal properties of such mesoscopic
structures may also lead to the discovery of new phenomena,
particularly when the quantum phase coherence length of the
thermal carriers is comparable to the sample dimensions.

The difficulty in making measurements of thermal properties on
mesoscopic devices stems from the problem of accurately measuring
the temperature on such a small size scale, without disturbing the
device being measured.  Although numerical estimates of thermal
properties of mesoscopic samples may be obtained by modeling the
heat flow, quantitative measurements have only recently become
possible.  Recently, in a beautiful experiment, Schwab et al.\cite{Schwab}
were able to measure the quantization of heat conduction in a
ballistic phonon waveguide using sophisticated lithographic and
measurement techniques.  However, equivalent
measurements of the thermal properties associated with electronic
conduction in mesoscopic metallic samples have not been reported.
In this Communication, we describe our quantitative measurements of
the electronic thermoelectric power and the thermal conductance of
single, doubly-connected, micron-size heterostructures formed from
a superconductor and a normal metal, using local thermometry
techniques that we recently developed \cite{aumentado}. Although the
thermal conductance shows no dependence on magnetic field to
within our measurement sensitivity, the thermopower shows
oscillations as a function of magnetic field, demonstrating the
phase coherent nature of thermal transport in this regime.

The electrical current $I$ and thermal current $I^T$ through a
metallic sample are related to the voltage difference $\Delta V$
and the temperature difference $\Delta T$ by the transport
equations \cite{ashcroft}:
\begin{equation}
I = G \Delta V + \eta \Delta T
\label{eq:eqn1}
\end{equation}
and
\begin{equation}
I^T = \zeta \Delta V + \kappa \Delta T
\label{eq:eqn2}
\end{equation}
Conceptually at least, thermal measurements on metallic samples
are relatively straightforward. A temperature differential $\Delta
T$ is applied to the sample, and the voltage $\Delta V$ across the
sample is measured, under the condition that the current $I$
through the sample is 0. The resulting ratio $S = {\Delta
V}/{\Delta T} = \eta/G$ is called the thermopower. The ratio $G^T
= I^T /\Delta T$ measured under the same conditions is the thermal
conductance. Replacing $\Delta V = S\Delta T$ from Eq. (1) into
Eq. (2), one obtains $G^T = S\zeta + \kappa$. For typical metals,
the first term in this expression is much smaller than the second
one, so that $G^T$ can be approximated by $G^T \approx \kappa$.

In order to obtain quantitative measurements of the thermal conductance and
thermopower on mesoscopic samples, one needs to accurately measure the local electron
temperature on submicron length scales. The requirement of any electron thermometer on
this size scale is that it should not disturb the local electron gas appreciably,
while still remaining sensitive to changes in the temperature. Recently, the Saclay
group has demonstrated in a series of beautiful experiments the possibility of using
normal-metal/insulator/superconductor (NIS) tunneling spectroscopy
to probe the local electron distribution in a mesoscopic
normal-metal sample \cite{pothier1}. While these thermometers can
accurately measure the electron temperature on the size
scale of approximately one hundred nanometers, the difficulty of
fabricating the tunnel junctions (usually done by \textit{in situ} shadow
evaporation techniques) precludes using this technique on more
complex sample geometries. In addition, determination of the
temperature requires measuring a full dc current voltage
characteristic at each point, and then fitting the measured curve
in order to obtain the temperature, making it time-consuming to
use when many data points are to be obtained. An alternative
technique that we have developed \cite{aumentado} is to use the strong
temperature dependence arising from the superconducting proximity
effect in a normal metal \cite{belzig} as a local electron temperature
thermometer. As we demonstrate below, the thermometers are
relatively simple to fabricate and measure, and provide the
ability to measure the local electron temperature at essentially
any point on a complex mesoscopic sample.

\begin{figure}[p]
\begin{center}
\BoxedEPSF{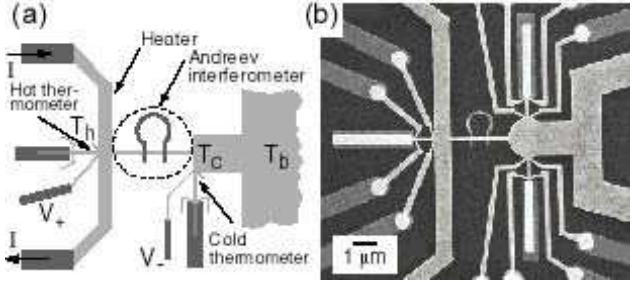 scaled 760}
\end{center}
\caption{(a) Schematic of our sample design for measuring the thermal
properties of a mesoscopic NS structure, an Andreev interferometer. Dark
gray is superconductor, lighter area is normal metal. The Andreev
interferometer is encircled by a dotted line.  
(b) Scanning electron micrograph of one of our actual devices fabricated by
conventional electron-beam lithography techniques.  Dark gray areas are
superconductor (Al), while the lighter areas are normal metal (Au).  For
thermopower measurements, the thermal voltage is measured using the contacts
labeled V$_+$ and V$_-$, as seen in (a).}
\end{figure}

Figure 1 shows a schematic and a scanning electron micrograph of
one of the devices we have measured. Three such samples were measured, although the results of
only one are presented here.  The devices were fabricated
by conventional multi-level electron-beam lithography technique on oxidized
Si substrates. The 65 nm thick Au wires and contacts were
patterned and evaporated first, after which the 65 nm thick Al film 
was evaporated following O$_2$ plasma etching to ensure good interfaces
between the Au and Al films. 
From weak localization measurements\cite{Mohanty} on long Au wires with
similar properties, we determined the electron phase coherence length to
be $L_{\phi}\cong$ 3.5 $\mu$m at $T$ = 300 mK, and the diffusion constant in the Au
to be 
$D \cong$ 1.9$\times 10^{-2}$ m$^2$/sec, resulting in a superconducting coherence
length in the normal metal $L_T\cong$ 0.38 $\mu$m at $T$=1 K. 

The sample itself is a so-called Andreev
interferometer \cite{nakano}, and consists of a normal metal Au wire
(2.44 $\mu$m long and 0.12 $\mu$m wide) interrupted  in the middle
by a superconductor (Al) to form a loop. Similar Andreev interferometers
have been recently shown to exhibit a number of interesting properties; in
particular, the resistance \cite{pothier2} and thermopower \cite{eom} show periodic
oscillations as a function of magnetic field, with a fundamental
period corresponding to a superconducting flux quantum $h/2e$
through the area of the loop. The thermometers on either end of
the Andreev interferometer consist of single Au wires with four
probes connected to Al leads. A superconducting film placed close
to each Au wire serves to induce a proximity effect \cite{belzig}. With
this design, simple heat flow simulations show that the
temperature profile across the thermometer is essentially flat
\cite{nagaev}.  At one end of the Andreev interferometer, a heater is
formed from a wide Au strip that is in direct electrical contact
with the Andreev interferometer and one thermometer (the `hot'
thermometer).  Two other thermometers measure the temperature at
the other end of the Andreev interferometer (the `cold' thermometers, only
one is used). By passing a dc current through the heater, one can heat one
end of the Andreev interferometer to a temperature above the substrate
temperature $T_b$.  Electrical  contact to the heater and all thermometers
is made through superconductors whose thermal conduction is small, so that at
temperatures below approximately half the critical temperature
$T_c$ of the Al, the power $P$ generated in the heater can flow
out essentially only through the Andreev interferometer. (The
electron-phonon scattering rate is at least three orders of magnitude
smaller than the electron-electron scattering rate at these
temperatures\cite{Wind}, so that heat flow through the phonons is
much smaller than the electronic conduction through the interferometer, 
and is ignored in our analysis.) The normal metal parts of the heater,
thermometers, and the Andreev interferometer itself are fabricated at the
same time, so that the coupling of the electrons in the thermometer with the
electrons in the heater and samples is very good.  By measuring the voltage
drop across the heater, and knowing the current through it, we therefore
have a quantitative estimate of the heat flow $I^T = P$ through the Andreev
interferometer.

The thermometers are first calibrated by measuring their
four-terminal resistance with an ac resistance bridge with no
current in the heater as a function of the temperature of the
cryostat, which in this experiment was a $^3$He sorption
refrigerator with a base temperature of 260 mK. The temperature of
the refrigerator is then kept fixed, and the ac resistance of the
thermometers is measured as a function of dc current through the
heater. By cross-correlation of the two measurements, one can
obtain the temperature of the electrons in the thermometers as a
function of the power through the heater. Fig. 2(a) shows the result
of these measurements for the hot and cold thermometer. At low
heater power ($\leq$ 10 pW), only the hot thermometer shows a
change in temperature, while at higher heater powers, both
thermometers show an increase in electron temperature. We also
show in the same figure the difference in temperature $\Delta T$
as a function of heater power, which grows as expected.

The thermal conductance of the Andreev interferometer can now be
essentially read directly from Fig. 2(a), since it is given by $G^T
= P/\Delta T$. Fig. 2(b) shows a plot of $P/\Delta T$ as a function
of the average temperature. Ideally, one should measure the
thermal conductance in the limit of $\Delta T \rightarrow 0$. This
limit can be obtained by extrapolating the curve in Fig. 2(b) to the
case when $T_{ave} = (T_h + T_c)/2$ approaches base
temperature, in the limit of zero power through the heater.
Performing this extrapolation, one obtains a conductance of $G^T
=$ 1.2$\times$10$^{-10}$ W/K.

\vspace{-.1cm}
\begin{figure}[p]
\begin{center}
\BoxedEPSF{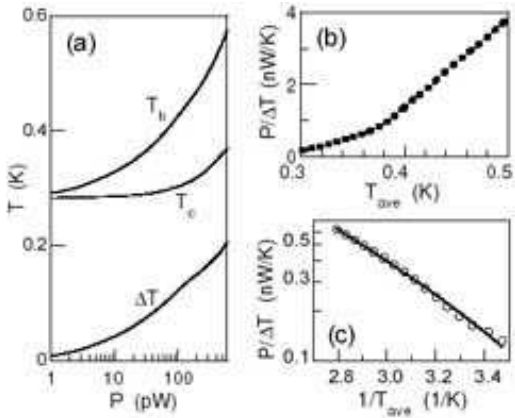 scaled 800}
\end{center}
\caption{(a) Temperature of hot and cold thermometers and their difference 
as a function of power dissipation in the heater at $T_b$ = 280 mK. 
(b) Thermal conductance as a function of the average temperature
$T_{ave}=(T_h+T_c)/2$ of the Andreev interferometer. 
(c) Low power regime in a semi-logarithmic plot, demonstrating the
exponential dependence of the conductance on inverse temperature. 
Solid line is a fit to the expected dependence for a superconductor.}
\end{figure}

In order to put this number in context, it is instructive to calculate the
thermal conductance of a gold wire of the same dimensions as in the Andreev
interferometer, but without the superconductor. This can be estimated from
the Wiedemann-Franz law \cite{ashcroft}, which states that the ratio of the thermal
to the electrical conductivity of a metal is proportional to the
temperature. In terms of conductances, one can write this as $G^T
 = AT/R$, where the textbook value \cite{ashcroft} of the constant $A$ for Au
is 2.32$\times$10$^{-8}$ W$\Omega$/K$^2$. With this value for $A$,
and the measured value of the normal state resistance $R$, we can
estimate the thermal conductance of the equivalent Au wire to be
1.3$\times$10$^{-9}$ W/K, more than an order of magnitude higher
than the measured thermal conductance of the Andreev
interferometer.

Although small deviations from the Wiedemann-Franz law are
expected for real metals, the order of magnitude difference
suggests that the superconductor present in the Andreev
interferometer has a substantial effect on its thermal
conductance. A reasonable first approximation is to estimate the
thermal conductance of the small section of the proximity coupled
normal part of Andreev interferometer (between the superconducting
arms) as a superconductor. The thermal conductance of a
superconductor arises solely from the presence of quasiparticles
in the superconductor, whose population is exponentially
suppressed at temperatures well below the superconducting gap.
Consequently, the thermal conductance of a superconductor at low
temperatures is given by an equation of the form\cite{abrikosov} 
\begin{equation}
G^{T}_S \approx
G^{T}_N \cdot\frac{6}{\pi^2}\left(\frac{\Delta}{k_{B}T}\right)^2
e^{ -\Delta/k_{B}T}, 
\label{eq:eqn3}
\end{equation}
where $G^{T}_N$ is the thermal conductance in the normal state, and
$\Delta$ is the superconducting gap. Fig. 2(c) shows a
plot of an expanded version of the low power regime of $G^{T}_S$
as a function of the inverse average temperature, along with a fit
to the equation above. From this, we obtain a gap $\Delta$ = 200
$\mu$eV, which compares favorably with the value $\Delta$ =183
$\mu$eV obtained from the measured $T_c$ of the Al.

As we noted earlier, the electrical conductance and thermopower of such
Andreev interferometers are expected to oscillate as a function of apllied
magnetic flux with a fundamental period $h/2e$, due to the quantum
interference of quasiparticles in the proximity coupled normal
metal\cite{belzig}. With the local thermometers we can now obtain
{\it quantitative} measurements of the thermopower oscillations. The 
thermal voltage across the Andreev interferometer is given by
\begin{equation}
\Delta V = \int_{T_c}^{T_{h}(I)}S_{A}dT \ , 
\label{eq:eqn4}
\end{equation}
where $S_A$ is the thermopower of the interferometer. We have
ignored here the thermopower contribution of the $V_+$ voltage
contact (see Fig. 1(a)), which is acceptable if it is small, or does
not vary as a function of external parameters (as is the case in
our experiments). In order to improve our sensitivity, we use an
ac technique \cite{eom} by superposing an ac tickling current on top
of the dc heater current, which implies that we measure the
derivative
\begin{equation}
\frac{d(\Delta V)}{dI} = S_A \frac{dT_h}{dI} \ ,
\label{eq:eqn5}
\end{equation}
where we have assumed that we are in the low current regime in
Fig. 2(a) ($\leq$ 10 $\mu$W), so that the temperature of the cold thermometer
is constant ($dT_c /dI$ = 0). $dT_h /dI$ can be numerically obtained
from $T_h (I)$, so that a quantitative estimate of the thermopower
can be directly obtained from the measured ac voltage.

\vspace{-.1cm}
\begin{figure}[p]
\begin{center}
\BoxedEPSF{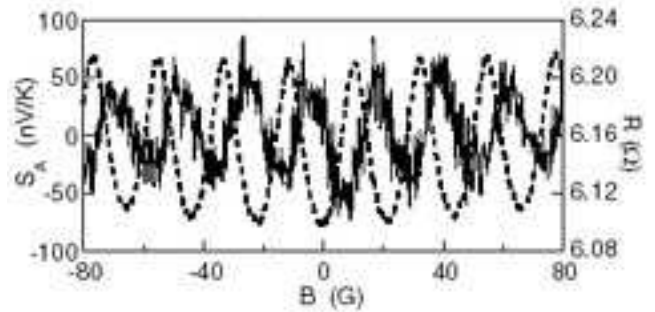 scaled 1000}
\end{center}
\caption{Magnetic flux dependence of the resistance (dashed curve) and
thermopower (solid curve) of the Andreev interferometer at $T_b$ = 295 mK.}
\end{figure}

Fig. 3 shows the thermopower $S_A$ and the resistance $R$ of the Andreev
interferometer as a function of the applied magnetic field $B$. The
thermopower data were taken with a dc current of 5 $\mu$A and an ac current
of rms amplitude 1 $\mu$A through the heater. Both quantities oscillate
as a function of $B$ with a fundamental period corresponding to a flux
$h/2e$ through the area of the interferometer. However, while the
oscillations in $R$ are symmetric in $B$, the oscillations in
$S_A$ are antisymmetric. A negative thermopower in a metal is
typically associated with electron-like charge carriers, while a
positive thermopower is associated with hole-like charge carriers.
An antisymmetric $S_A$ means that the modulation of the quantum
interference in the Andreev interferometer by the magnetic field
periodically changes the sign of the thermopower in the interferometer. 

Two aspects of our results are worthy of note. First, at the level of our
measurement sensitivity, the thermal conductance of the Andreev
interferometer does not oscillate with $B$, although one might expect such
oscillations based on the fact that $R$ and $S$ oscillate with $B$\cite{belzig}. 
Unlike the electrical properties, however, the thermal conductance of the
Andreev interferometer is determined by the series addition of the thermal
conductance of the proximity coupled normal-metal wires, and the small
section of the superconductor which lies across the normal metal. Since the
thermal conductance of this last section of the superconductor is about an
order of magnitude smaller than that of the normal metal regions, it
determines the thermal conductance of the entire sample. Consequently, the
small variations of the thermal conductance in the normal metal regions
associated with the proximity effect will not be observable. Second, the
symmetry of the thermpower with respect to magnetic field is similar to what
has been observed earlier\cite{eom}, where it was noted that this symmetry
(symmetric or antisymmetric with respect to field) appeared to depend on
sample geometry. The origin of this behavior is still not understood. The
measurements reported here confirm the results of the earlier experiments,
and may also provide a possible clue to the origin of the dependence of the
symmetry on sample topology. For Andreev interferometers where the
superconductor lies in the path of the heat current, the low thermal
conductivity of the superconductor implies that essentially all the
temperature differential is dropped across it, and none across the proximity
coupled normal metal, so that no thermal voltage is developed across the
normal metal (the superconductor, of course, has no thermal voltage). For
Andreev interferometers where the superconductor is not in the path of the
heat current, one would have a large contribution to the thermopower from
the proximity coupled normal metal. While this does not explain the symmetry
of the thermopower with respect to magnetic field, it does indicate why one
might expect zero thermopower contribution in the absence of a magnetic
field for Andreev interferometers of the first type, as is observed. Further
work is required to fully understand the magnetic field dependence.

This work was supported by the NSF through DMR-9801982, and by the
David and Lucile Packard Foundation.

\end{multicols}

\end{document}